\documentstyle[pra,aps,tighten]{revtex}

\begin{document}
\title{Quantum authentication using entangled state}
\author{Yong-Sheng Zhang, Chuan-Feng Li\thanks{%
Electronic address: cfli@ustc.edu.cn}, Guang-Can Guo\thanks{%
Electronic address: gcguo@ustc.edu.cn}}
\address{Laboratory of Quantum Communication and Quantum Computation, and Department\\
of Physics, University of Science and Technology of China, Hefei 230026, \\
People's Republic of China}
\maketitle

\begin{abstract}
A scheme of quantum authentication is presented. Two parties share
Einstein-Podolsky-Rosen (EPR) pairs previously as the authentication key
which servers as encoder and decoder. The authentication is accomplished
with local controlled-NOT operations and unitary rotations. It is shown that
our scheme is secure even in the presence of an eavesdropper who has
complete control over both classical and quantum channels. Another character
of this protocol is that the EPR sources are reusable. The robustness of
this protocol is also discussed.

PACS number(s): 03.67.Dd, 03.65.Bz\ 
\end{abstract}

\section{Introduction}

Quantum cryptography is a field which combines quantum theory with
information theory. The goal of this field is to use the laws of physics to
provide secure information exchange, in contrast to classical methods based
on (unproven) complexity assumption. Since the publication of BB84 protocol 
\cite{Ben84}, quantum key distribution (QKD) \cite{Ben84,Ekt91} has
developed into a well understood application of quantum mechanics to
cryptography. In particular, QKD protocols became especially important due
to technological advances which allow their implementation in the laboratory 
\cite{Jenn00}. To guarantee the security of the quantum key in practical
applications, quantum key authentication is important as well as QKD.
Moreover, large quantity of communication tasks in modern society need more
reliable authentication systems. However, up to now, the security of
practically used authentication systems is based on the computational
difficulty, i.e., they rely on limited advancement of computer power,
technologies, and mathematical algorithms in the foreseeable future.

Recently, several quantum authentication schemes \cite
{Duk99,GZ00,Bar99,Jen00} have been proposed. The scheme proposed by Du\v sek 
{\it et al.} \cite{Duk99} is a combination of classical identification
procedure and quantum key distribution. Zeng's \cite{GZ00} protocol is using
EPR pairs as the first authentication keys, and using classical keys
distributed in quantum key distribution procedure after then. Moreover, two
very interesting authentication protocols using entanglement and catalysis 
\cite{JP99} were proposed by Barnum \cite{Bar99} and Jensen {\it et al}{\bf .%
} \cite{Jen00} respectively. In their protocol, the two parties in
communication have previously shared catalyst (a particular pair of
entangled particles). The verifier sends the challenge which is a half part
of some entangled state to the identifier, then they can transfer this
entangled state to a special state (the state of the catalyst)
deterministically by local quantum operation and classical communication
(LQCC) with the catalyst. However, this task can not be accomplished by LQCC
without the catalyst. So the verifier can authenticate the identifier by
measuring the state of the challenge after the identifier sends it back.

On the other hand, entanglement of multiparticle system is a important
feature of quantum mechanics. In addition to their central role in
discussion of nonlocal quantum correlations, they form the basis of quantum
information such as quantum teleportation \cite{Ben93}, quantum key
distribution, quantum dense coding \cite{Ben92} {\it et al}. EPR pairs are
used as communication channels in protocols mentioned above. In this paper,
we present a authentication protocol with EPR state as the key (encoder and
decoder). The authentication is accomplished with local controlled-NOT
(C-NOT) operations and unitary rotations.

The paper is structured as follows. In Section II, we give the framework of
our authentication protocol, and security of this protocol is analyzed in
Section III. In Section IV, we consider the nonideal situation and discuss
the robustness of this protocol. Section V concludes the paper.

\section{Quantum authentication scheme}

The general task of authentication is verifying the identification of each
other of two parties (Alice and Bob) in communication, using quantum and
classical channel. The protocols are such that if Alice and Bob can
successfully complete one, Alice is convinced that Bob (or someone who has
stolen his identification token) is on the other end of the quantum
communication channel. The classical analogue of this can be done by having
Bob to reveal, over a classical channel, a secret which Alice and Bob had
previously securely shared. The quantum protocol presented here uses shared
entangled states as the counterpart of shared secret key. There is
protection, via the no-cloning theorem, against copying of the
authentication token. And this protocol may provide reusable authentication
tokens.

Alice and Bob have previously shared $2K$ pairs entangled states in 
\begin{equation}
\left| \Phi ^{+}\right\rangle =\frac 1{\sqrt{2}}\left( \left|
00\right\rangle +\left| 11\right\rangle \right) .  \eqnum{1}
\end{equation}
An authentication round consists of the following steps. First, for example,
Alice acts as identifier and Bob acts as verifier.

When the authentication begin, the two parties rotate their particles' state
by $\theta $ respectively. The rotation can be described as 
\begin{equation}
R(\theta )=\left( 
\begin{array}{cc}
\cos \theta & \sin \theta \\ 
-\sin \theta & \cos \theta
\end{array}
\right) .  \eqnum{2}
\end{equation}
The state $\left| \Phi ^{+}\right\rangle $ does not change under bilateral
operation of $R(\theta )$. The purpose of this operation is to prevent the
eavesdropper's impersonation. (The detailed interpretation will be given in
Section III.) Then Bob prepares $K^{\prime }$ ($K^{\prime }\leq K$)
particles $\gamma _B^i$ in arbitrary pure state 
\begin{equation}
\left| \psi _i\right\rangle =a_i\left| 0\right\rangle +b_i\left|
1\right\rangle ,\text{ }i=1,3,5,\cdots ,2K^{\prime }-1,  \eqnum{3}
\end{equation}
where $\left| a_i\right| ^2+\left| b_i\right| ^2=1$, $a_i$ and $b_i$ are two
complex numbers which are selected randomly. The state $\left| \psi
_i\right\rangle $ is only known by Bob himself. Bob sends the challenge
particle $\gamma _B^i$ to Alice in order of $i$ (where $i$ is odd). Thus
Alice uses the corresponding particle $\beta _A^i$ of the entangled pairs
and the particle $\gamma _B^i$ to do a C-NOT operation ($\beta _A^i$ is
controller and $\gamma _B^i$ is target) and the three particles' state will
be 
\begin{equation}
\left| \Psi _i\right\rangle =\frac 1{\sqrt{2}}\left( a_i\left|
000\right\rangle +b_i\left| 001\right\rangle +a_i\left| 111\right\rangle
+b_i\left| 110\right\rangle \right) ,  \eqnum{4}
\end{equation}
then she sends back $\gamma _B^i$ to Bob. Bob uses his corresponding
particles $\beta _B^i$ (which entangled with $\beta _A^i$) to do a C-NOT
operation on $\gamma _B^i$ again. Now the state of the key particles and the
challenge $\gamma _B^i$ is the same as it at the first 
\begin{equation}
\left| \Psi _i^{\prime }\right\rangle =\frac 1{\sqrt{2}}\left( \left|
00\right\rangle +\left| 11\right\rangle \right) \otimes \left( a_i\left|
0\right\rangle +b_i\left| 1\right\rangle \right) =\left| \Phi
^{+}\right\rangle \otimes \left| \psi _i\right\rangle .  \eqnum{5}
\end{equation}
Bob measures $\gamma _B^i$ in basis $\left| \psi _i\right\rangle $ and $%
\left| \psi _i\right\rangle ^{\perp }$ (state orthogonal to $\left| \psi
_i\right\rangle $). If $\gamma _B^i$ is in state $\left| \psi
_i\right\rangle $, it passes the test; otherwise, it fails and Bob aborts
the protocol. Then Alice becomes the verifier, she prepares $K^{\prime }$
challenge $\gamma _A^i$ and uses her particle $\beta _A^i$ ($i$ is even) to
do the same steps.

The authentication fails if any of the projective measurements in the
previous step fails, or if Alice or Bob receive more than $K^{\prime }$
requests to send back challenge particles.

If the authentication round succeeds, Alice and Bob retain all $2K$ pairs of
entangled states and can reuse them in the next time. However, the security
of the entangled states used later is a little less than the original one.
If the authentication fails, the parties discard all particles used till
that point. In this case, Alice and Bob have to start again with new keys
(EPR pairs).

\section{Security analysis}

We now discuss the security of this protocol. First, Eve may impersonate
Alice when Alice is not present. When Bob sends out a challenge $\gamma _B^i$%
, Eve intercepts it which she can manipulate using unitary transformation or
measurement. Since Eve has not shared the key with Bob, she can not entangle
this challenge with Bob's key particles. Suppose the state Eve send back is 
\begin{eqnarray}
\rho _i &=&\sum_{k=1}^2p_{ik}\left| \psi _{ik}^{\prime }\right\rangle
\left\langle \psi _{ik}^{\prime }\right| ,  \eqnum{6} \\
\left| \psi _{ik}^{\prime }\right\rangle &=&a_{ik}^{\prime }\left|
0\right\rangle +b_{ik}^{\prime }\left| 1\right\rangle .  \nonumber
\end{eqnarray}
where $p_{i1}+p_{i2}=1$ and $\left| a_{ik}^{\prime }\right| ^2+\left|
b_{ik}^{\prime }\right| ^2=1$. After Bob's C-NOT operation, the fidelity 
\cite{Joz94} of the state and the test state $\left| \Psi _i\right\rangle $
is 
\begin{equation}
F=\frac 12+\sum_{k=1}^2p_{ik}\left[ Re\left( a_i^{*}b_ia_{ik}^{\prime
}b_{ik}^{\prime *}\right) +Re\left( a_i^{*}b_ia_{ik}^{\prime
*}b_{ik}^{\prime }\right) \right] ,  \eqnum{7}
\end{equation}
where $Re(x)$ is the real part of complex number $x$. Since $a_i$ and $b_i$
are selected randomly, the average value of $F$ is $\frac 12$. That is to
say Eve has the probability only $\frac 12$ to pass the test in one time on
the average. The probability that Eve is not detected by Bob is $\left( 
\frac 12\right) ^{K^{\prime }}$ on the average, which can be made
arbitrarily small by choosing $K^{\prime }$ large enough.

Second, Eve may use the method of denial of service. In this type of attack,
Eve deliberately causes the authentication round to fail, and hence causes
one party to discard all key particles. Although this protocol is
particularly vulnerable to this kind of attack, this is not an essential
weakness, since an attacker who controls both quantum and classical
communication can always prevent successful authentication between the
legitimate parties.

We now look at stronger attacks in which Eve tries to obtain key material
which she could then use, e.g., in a later impersonation attack. Eve's goal
is to share pairs of particles in the entangled state $\left| \Phi
^{+}\right\rangle $ with Alice and/or Bob. For instance, if she succeeds in
obtaining a large amount of key material with Bob, she will be able to
authenticate herself to Bob without Alice being present. However, if Eve's
presence is detected in a single measurement, all the previously obtained
key material she shares with the verifier who performed that measurement
will be worthless.

Eve has three choices to attack: (I) She can intercept the challenge Bob
sends to Alice, and make her own particle interact with this challenge, then
send the challenge or her own particles to Alice. (II) She can pass the
challenge to Alice but intercept it when Alice sends it back to Bob, and
make her own particles interact with this challenge. (III) She can combine
the strategies (I) and (II). We will analyze Eve's three strategies
respectively.

(I) Since the challenge Bob sends to Alice is not entangled with the key
particles, Eve can not share key material with Bob or Alice. If she changes
the state of the challenge, she will be detected with probability $1-F$, and
the form of $F$ is the same as Eq. (6).

(II) In this case, Eve may make his own particles entangle with Alice and
Bob's key pairs. However, the challenge state is selected randomly, Eve can
not make her own particle be maximally entangled with Bob and/or Alice's
particle deterministically, and the probability she will be detected is $1-F$
too.

However, there exist other more powerful strategy.

(III) Eve intercept the challenge Bob send to Alice, and send her own
particle in state $\left| 0\right\rangle $ or $\left| 1\right\rangle $ to
Alice. After Alice's operation, this particle will be entangled with Alice
and Bob's key particles in GHZ state, and Eve can use the key to complete
authentication with Alice or Bob as efficient as the key shared by Alice and
Bob. We can defend this attack by an additional step at first. Before Bob
sends the challenge to Alice, they do a bilateral rotation $R(\theta
_i)=\left( 
\begin{array}{cc}
\cos \theta _i & \sin \theta _i \\ 
-\sin \theta _i & \cos \theta _i
\end{array}
\right) $. The state of the maximally entangled two particles will be
unchanged when both Alice and Bob rotate the $i$th particle by $\theta _i$.
However if Eve has entangled her particle with Alice and Bob's particles in
state $\left| \Phi \right\rangle _{ABE}=\frac 1{\sqrt{2}}\left( \left|
000\right\rangle +\left| 111\right\rangle \right) $. In the second
authentication process between Alice and Bob, the state will be changed to 
\begin{eqnarray}
\left| \Phi _i\right\rangle _{ABE} &=&\cos ^2\theta _i\frac 1{\sqrt{2}}%
\left( \left| 000\right\rangle +\left| 111\right\rangle \right) +\sin
^2\theta _i\frac 1{\sqrt{2}}\left( \left| 110\right\rangle +\left|
001\right\rangle \right)  \eqnum{8} \\
&&+\sin \theta _i\cos \theta _i\frac 1{\sqrt{2}}\left( \left|
011\right\rangle -\left| 100\right\rangle \right) +\sin \theta _i\cos \theta
_i\frac 1{\sqrt{2}}\left( \left| 101\right\rangle -\left| 010\right\rangle
\right)  \nonumber
\end{eqnarray}
under the bilateral rotation. When Bob or Alice measures the challenge, the
average fidelity will be $1-\frac 12\sin ^2\theta _i$, and the average
probability that they find Eve's attack will be $\frac 12\sin ^2\theta _i$.
If $\theta _i$ is selected randomly and known only by Alice and Bob (it can
be previously shared and reusable), the average fidelity of the challenge
when Bob test it is $\frac 34$ and the average probability that Eve will be
detected is $\frac 14$. If Eve tries to impersonate Alice or Bob with the
entangled state $\left| \Phi \right\rangle _{ABE}$, it is easy to verify
that the average probability for Eve to succeed is less than $1-\frac 12\sin
^2\theta _i$ (even Eve knows exactly the value of $\theta _i$).

Now we consider to use a fixed rotation angle $\theta $ for all
authentication rounds.

There exist two special angle, that Eve has two corresponding strategies to
impersonate or to obtain the authentication key as her particle has been
already entangled with Alice and Bob's key particles in GHZ state. When $%
\theta =0$ (or $\frac \pi 2$, $\pi $ {\it et al}.), Eve can use her
particles impersonate Alice without risk that Bob will detect the
impersonation. Another special angle is $\theta =\frac \pi 4$ (or $\frac{%
3\pi }4$, $\frac{5\pi }4$ {\it et al}.). In this case, after Alice and Bob
rotate their particle by $\frac \pi 4$, Eve can rotate her particle by $%
\frac \pi 4$ too. Then Eve intercepts the challenge Bob sends to Alice and
sends her own particle to Alice, after Alice does a C-NOT operation on Eve's
particle, it will be entangled with Bob's particle in EPR state and Alice's
particle has no quantum correlation with Bob and Eve's particles. The
process can be described as 
\begin{eqnarray}
\ \ \left| \Phi \right\rangle _{ABE} &=&\frac 1{\sqrt{2}}\left[ \left|
000\right\rangle +\left| 111\right\rangle \right] _{ABE}  \eqnum{9} \\
&&\ \stackrel{R}{\rightarrow }\ \ \frac 14\left[ \left( \left|
0\right\rangle -\left| 1\right\rangle \right) \left( \left| 0\right\rangle
-\left| 1\right\rangle \right) \left( \left| 0\right\rangle -\left|
1\right\rangle \right) +\left( \left| 0\right\rangle +\left| 1\right\rangle
\right) \left( \left| 0\right\rangle +\left| 1\right\rangle \right) \left(
\left| 0\right\rangle +\left| 1\right\rangle \right) \right] _{ABE} 
\nonumber \\
&&\ \stackrel{C_{AE}}{\rightarrow }\frac 12\left( \left| 0\right\rangle
+\left| 1\right\rangle \right) _A\otimes \left( \left| 00\right\rangle
+\left| 11\right\rangle \right) _{BE},  \nonumber
\end{eqnarray}
where $R=R_A\left( \pi /4\right) \otimes R_B\left( \pi /4\right) \otimes
R_E\left( \pi /4\right) $ and $C_{AE}$ is a C-NOT operation (A controls E).
After Alice sent Eve's particle back, Eve uses it to do a C-NOT operation on
Bob's challenge, then sends the challenge back to Bob for verification. Up
to now, Eve has obtained the authentication key without disturbing the
authentication process between Alice and Bob.

For an arbitrary $\theta $ (we assume that Eve knows this angle), if Eve
tries to impersonate, in the case of $\sin ^2\theta <\cos ^2\theta $, the
optimal strategy for Eve is to do a C-NOT operation on the challenge
directly and send back it. The probability she will succeed is $1-\frac 12%
\sin ^2\theta $. In the case of $\sin ^2\theta >\cos ^2\theta $, Eve should
do a NOT operation on her particle first, then do a C-NOT operation on the
challenge before she sends it back. The probability she will succeed is $1-%
\frac 12\cos ^2\theta $. Sum the two cases, the probability Eve will succeed
is 
\begin{equation}
P_1=%
\mathop{\rm max}%
\left\{ 1-\frac 12\sin ^2\theta ,1-\frac 12\cos ^2\theta \right\} . 
\eqnum{10}
\end{equation}

If Eve tries to get the key, she does a rotation on her particle, then sends
her particle to Alice to do a C-NOT operation on the particle and Eve can do
a another proper rotation on her particle when Alice sends it back 
\begin{eqnarray}
\left| \Phi \right\rangle _{ABE} &=&\frac 1{\sqrt{2}}\left[ \left|
000\right\rangle +\left| 111\right\rangle \right] _{ABE}\stackrel{R_{ABE}}{%
\rightarrow }\ \left| \Phi \left( \theta ,\phi _1\right) \right\rangle _{ABE}
\eqnum{11} \\
&&\stackrel{C_{AE}}{\rightarrow }\left| \Phi ^{\prime }\left( \theta ,\phi
_1\right) \right\rangle _{ABE}\stackrel{R_E}{\rightarrow }\left| \Phi \left(
\theta ,\phi _1,\phi _2\right) \right\rangle _{ABE},  \nonumber
\end{eqnarray}
where $R_{ABE}=R_A\left( \theta \right) \otimes R_B\left( \theta \right)
\otimes R_E\left( \phi _1\right) $ and $R_E$ is operation $R_E\left( \phi
_2\right) $. Then the probability $P_2$ that Eve will succeed to obtain the
key is 
\begin{eqnarray}
P_2 &=&\max_{\phi _1,\phi _2}\left\{ F\left( \left| \Phi ^{+}\right\rangle
\left\langle \Phi ^{+}\right| ,tr_A\left( \left| \Phi \left( \theta ,\phi
_1,\phi _2\right) \right\rangle _{ABE}\left\langle \Phi \left( \theta ,\phi
_1,\phi _2\right) \right| _{ABE}\right) \right) \right\}  \eqnum{12} \\
\ &=&\frac 12\left( \left| \cos \theta \right| +\left| \sin \theta \right|
\right) ^2.  \nonumber
\end{eqnarray}

Now we consider the optimal angle Alice and Bob should select if they use a
fixed angle $\theta $ for all EPR pairs. Using the principle $P_1=P_2$, we
can get 
\begin{equation}
\left| \cos \theta \right| =\frac 2{\sqrt{5}}\text{ },\frac 1{\sqrt{5}};%
\text{ }P_1=P_2=\frac 9{10}.  \eqnum{13}
\end{equation}
From above we can know, if the authentication process use a fixed rotation
angle $\theta $, the successful probability of Eve's eavesdropping will be
increased.

\section{Robustness of the protocol}

Up to this point, our discussion has assumed that the initial state is ideal
maximal entangled state $\left| \Phi ^{+}\right\rangle $. Suppose, however,
that this state is corrupted a little after reused for many times, Alice and
Bob have a state described by density matrix 
\begin{equation}
\rho =\left( 1-\epsilon \right) \left| \Phi ^{+}\right\rangle \left\langle
\Phi ^{+}\right| +\epsilon \rho _1,  \eqnum{14}
\end{equation}
where $\epsilon $ is a parameter of the deviation of $\rho $ from $\left|
\Phi ^{+}\right\rangle \left\langle \Phi ^{+}\right| $ and $\rho _1$ is an
arbitrary state. Our results are most easily presented using the {\it trace
distance}, a metric on Hermitian operators defined by $T\left( A,B\right)
=Tr\left( \left| A-B\right| \right) $ \cite{Vi99}, where $\left| X\right| $
denotes the positive square root of the Hermitian matrix $X^2$. From above,
we can get that $T\left( \left| \Phi ^{+}\right\rangle \left\langle \Phi
^{+}\right| ,\rho \right) \leq 2\sqrt{\epsilon }$.

Ruskai \cite{Ru94} has shown that the {\it trace distance} contracts under
physical processes. If all operations are exact in the next authentication
process, since the state $\left| \Phi ^{+}\right\rangle $ will be unchanged
in the process, the density matrix $\rho $ will be transformed to 
\begin{equation}
\rho ^{\prime }=\left( 1-\epsilon \right) \left| \Phi ^{+}\right\rangle
\left\langle \Phi ^{+}\right| +\epsilon \rho _1^{\prime },  \eqnum{15}
\end{equation}
and $T\left( \left| \Phi ^{+}\right\rangle \left\langle \Phi ^{+}\right|
,\rho ^{\prime }\right) \leq 2\sqrt{\epsilon }$.

The fidelities $F\left( \left| \Phi ^{+}\right\rangle \left\langle \Phi
^{+}\right| ,\rho \right) $ and $F\left( \left| \Phi ^{+}\right\rangle
\left\langle \Phi ^{+}\right| ,\rho ^{\prime }\right) $ are both no less
than $1-\epsilon $, so the probability that the authentication fails is no
more than $\epsilon $. In conclusion, we can say that this protocol is
robust.

\section{Conclusion}

A scheme of quantum authentication using entangled state is presented. Two
parties share EPR pairs previously as the authentication key which servers
as encoder and decoder. The authentication is accomplished with local
controlled-NOT operations and unitary rotations. This protocol appears to be
secure even in the presence of an eavesdropper who has complete control over
both classical and quantum communication at all times. Our protocol is not
rely on classical cryptography, and needs not communication of classical
information of measurement results or details of operation method except the
indices of the particle Bob sends to Alice. Compared to the scheme using
catalysis, this protocol uses state only two dimensions, instead of five
dimensions. And compare to other authentication schemes, this protocol has
the same advantage that the keys are reusable as scheme using catalysis. At
the end, we expect that this method can be applied in quantum key
distribution (QKD) \cite{Zha00}.

\begin{center}
{\bf Acknowledgments}
\end{center}

We would like to thank Zheng-Wei Zhou for the very helpful discussions. This
work was supported by the National Natural Science Foundation of China.\


\begin{references}
\bibitem{Ben84}  C. H. Bennett and G. Brassard, in{\it \ Proceedings of the
IEEE International Conference on Computers, Systems, and Signal Processing,
Bangalore, India, }1984 (IEEE, New York, 1984), p. 175.

\bibitem{Ekt91}  A. K. Ekert, Phys. Rev. Lett. {\bf 67}, 661 (1991).

\bibitem{Jenn00}  T. Jennewein, C. Simon, G. Weihs, H. Weinfurter, and A.
Zeilinger, Phys. Rev. Lett. {\bf 84}, 4729 (2000); D. S. Naik, C. G.
Peterson, A. G. White, A. J. Berglund, and P. G. Kwiat, Phys. Rev. Lett. 
{\bf 84}, 4733 (2000); W. Tittel, J. Brendel, H. Zbinden, and N. Gisin,
Phys. Rev. Lett. {\bf 84}, 4737 (2000).

\bibitem{Duk99}  M. Du\v sek, O. Haderka, M. Hendrych, and R. My\v ska,
Phys. Rev. A {\bf 60}, 149 (1999).

\bibitem{GZ00}  G. Zeng and W. Zhang, Phys. Rev. A {\bf 61}, 022303 (2000).

\bibitem{Bar99}  H. Barnum, e-print arXiv: quant-ph/9910072.

\bibitem{Jen00}  J. G. Jensen and R. Schack, e-print arXiv: quant-ph/0003104.

\bibitem{JP99}  D. Jonathan and M. Plenio, Phys. Rev. Lett. {\bf 83}, 3566
(1999).

\bibitem{Ben93}  C. H. Bennett, G. Brassard, C. Cr\'epeau, R. Jozsa, A.
Peres and W. K. Wootters, Phys. Rev. Lett. {\bf 70}, 1895 (1993).

\bibitem{Ben92}  C. H. Bennett and S. J. Wiesner, Phys. Rev. Lett. {\bf 69},
2881 (1992).

\bibitem{Joz94}  R. Jozsa, J. Mod. Opt. {\bf 41}, 2315 (1994).

\bibitem{Ru94}  M. B. Ruskai, Rev. Math. Phys. {\bf 6}, 1147 (1994).

\bibitem{Vi99}  G. Vidal, D. Jonathan and M. A. Nielsen, e-print arXiv:
quant-ph/9910099.

\bibitem{Zha00}  Y.-S. Zhang, C.-F. Li, G.-C. Guo, to be submitted.
\end{references}
\end{document}